\documentclass{mem}
\usepackage{natbib}\usepackage{txfonts}\usepackage{balance}
\usepackage{graphicx}
\usepackage[a4paper,breaklinks,dvipdfm]{hyperref}
\idline{1}{1}
\begin{document}
\def\teff{$T\rm_{eff }$}
\def\kms{$\mathrm {km s}^{-1}$}

\title{
Cosmic Rays and Radiative Instabilities
}

   \subtitle{}

\author{
T. W. \,Hartquist\inst{1}
\and A. Y. \,Wagner\inst{2}
\and S. A. E. G. \,Falle\inst{3}
\and J. M. Pittard\inst{1}
\and S. Van Loo\inst{4,5}}

    \offprints{T. W. Hartquist}

\institute{
School of Physics and Astronomy, University of Leeds, Leeds LS2 9JT, UK
\and Center for Computational Sciences, Tsukuba University, 1-1-1 Tennodai,
Tsukuba Ibaraki, Japan 305-8577
\and Department of Applied Mathematics, University of Leeds, Leeds LS2
9JT, UK
\and Department of Astronomy, University of Florida, Gainesville, Florida
32611, USA
\and Harvard-Smithsonian Center for Astrophysics, 60 Garden Street,
Cambridge, Masschusetts 02138, USA  \email{t.w.hartquist@leeds.ac.uk}
}

\authorrunning{Hartquist et al. }

\titlerunning{Cosmic Rays and Radiative Instabilities}

\abstract{In the absence of magnetic fields and cosmic rays, radiative
cooling laws with a range of dependences on temperature affect the stability
of interstellar gas. For about four and a half decades, astrophysicists
have recognised the importance of the thermal instablity for the
formation of clouds in the interstellar medium. Even in the past several
years, many papers have concerned the role of the thermal instability
in the production of molecular clouds. About three and a half decades
ago, astrophysicists investigating radiative shocks noticed that for many
cooling laws such shocks are unstable. Attempts to address the effects
of cosmic rays on the stablity of radiative media that are initially
uniform or that have just passed through shocks have been made. The
simplest approach to such studies involves the assumption that the cosmic
rays behave as a fluid. Work based on such an approach is described.
Cosmic rays have no effect on the stability of initially uniform, static
media with respect to isobaric perturbations, though they do affect
the stability of such media with respect to isentropic perturbations.
The effect of cosmic rays on the stability of radiative shocked media
depends greatly on the efficiency of the conversion of energy in accelerated
cosmic rays into thermal energy in the thermalized fluid. If that efficiency
is low, radiative cooling makes weak shocks propagating into
upstream media with low cosmic-ray pressures more likely to be cosmic-ray
dominated than adiabatic shocks of comparable strength. The cosmic-ray
dominated shocks do not display radiative overstability. Highly efficient
conversion of cosmic-ray energy into thermal energy leads shocked media
to behave as they do when cosmic rays are absent.

\keywords{shock waves - - ISM: kinematics and dynamics - - ISM: cosmic rays -
- hydrodynamics - - instabilities - - ISM: clouds}
}
\maketitle{}

\section{Introduction}

After briefly reviewing the thermal instability of a non-magnetized,
static, uniform fluid containing no cosmic rays, we mention the importance
of the thermal instability in recent models of molecular cloud formation and
then consider the effects of cosmic rays on the instability of a medium
like that described above. Then we describe work on the radiative
instability of non-magnetized shocked fluids containing no cosmic rays,
before addressing the effects of cosmic rays on such media.

\section{Thermal Stability of a Uniform Fluid with No Cosmic Rays}

In a static equilibrium fluid containing no magnetic field or cosmic
rays

\begin{eqnarray}
&\rho\mathcal{L}(\rho,T)=0
\end{eqnarray}

\noindent $\rho$ and $T$ are the mass density and temperature, respectively,
and $\rho\mathcal{L}$ is the net energy loss per unit volume per unit time. 
In order to write the criteria for thermal stability \citep{Field:1965} of such a
state, we introduce some notation.

\begin{eqnarray}
%&z\equiv\frac{\omega}{ak}\\
&a\equiv\left(\frac{\gamma_gP_{g0}}{\rho_0}\right)^{\frac{1}{2}}\\
&k_T\equiv\frac{\mu(\gamma_g-1)}{Ra}\left.\frac{\partial\mathcal{L}}{\partial{}T}\right|_{T=T_0}\\
&k_\rho\equiv\frac{\mu(\gamma_g-1)\rho_0}{RaT_0}\left.\frac{\partial\mathcal{L}}{\partial\rho}\right|_{\rho=\rho_0}
%&k_c\equiv\frac{a}{\chi}\\
%&\phi\equiv\frac{\gamma_cP_{c0}}{\gamma_gP_{g0}}.
\end{eqnarray}

\noindent $\gamma_g$ is 5/3 for a fully ionized gas, $R$ is the gas
constant, $\mu$ is the mean mass per particle, and the subscript zero
indicates that the value of the quantity is that appropriate for the
equilibrium state.

Such a medium is stable with respect to isobaric perturbations if

\begin{eqnarray}
&k_T-k_\rho>0
\end{eqnarray}

\noindent It is stable with respect to isentropic perturbations if

\begin{eqnarray}
&k_T-\Delta>0 \\
&\Delta\equiv\frac{k_T-k_\rho}{\gamma_g}
\end{eqnarray}

Often $\mathcal{L}$ can be written as $\Lambda(T)n - \Gamma$, where $n$ is
the number density of nuclei, $\Lambda$ depends on only the temperature,
and $\Gamma$ is a constant.
When this is appropriate, stability with respect to isobaric perturbations
requires that $\alpha \equiv (T/\Lambda)(d\Lambda/dT)$ exceeds unity and
stability with respect to isentropic perturbations requires that $\alpha$
exceeds $-1.5$. 

\section{The Significance of Thermal Instability for Cloud Formation}

For conditions approaching the so-called typical interstellar conditions,
the function $n(P_g)$, satisfying $\mathcal{L}(n, P_g) = 0$, is single-valued
for low values of $P_g$ \citep[e.g.,][]{Wolfire:1995}; the corresponding value of $T$ is
of the order of
$10^4$K. For a range of values of $P_g$ within a factor of a few of $4 \times
10^{-13}$ erg cm$^{-3}$, $n$ is treble-valued. The state corresponding to
the intermediate solution is thermally unstable, and the temperatures of the
other two states are of the order of $10^4$K and $10^2$K, respectively. For
high values of $P_g$, $n$ is single-valued, and the corresponding value of
$T$ is of the order of $10^2$K.

The structure of the $n(P_g)$ curve implies that a shock of fairly modest
strength, encountering a medium that initially is cloudless and has $P_g$
only a bit below the typical interstellar value, can increase the pressure
enough that some of the gas becomes thermally unstable and forms clouds.
Such pictures for the formation of molecular clouds have been explored
for over a decade. At first, much work was focussed on non-magnetic models.
The many recent models of collisions of magnetized streams of
gas that trigger cloud formation due to thermal instability include those
of \citet{Hennebelle:2008} and \citet{Heitsch:2009}.

Models of shocks interacting with radiative, magnetized clumps of gas
leading to the formation of dense clouds include the two-dimensional
models of \citet{Lim:2005} and \citet{VanLoo:2007} and the
three-dimensional model of \citet{VanLoo:2010}. The finite extent of
the obstacles alleviates the over efficiency of star formation
associated with the colliding stream models. However, collisions between
clumps of finite extent may trigger cloud and star formation
at least as frequently as supernova remnant or superbubble
shocks encountering clumps. In any case, \citet{VanLoo:2007} found
considerable similarities between the structure of their model and
the observed structure of W3. They also illuminated the different
roles of the fast-mode and slow-mode shocks. A fast-mode shock propagates
into the clump first, and behind it radiative losses create magnetically
dominated regions. The following slow-mode shock compresses the gas and
creates a dense shell in which the magnetic pressure is not much larger
than the thermal pressure. Van Loo and his collaborators
have shown that the triggering shock must be of moderate strength. Shocks
with Alfv{\'e}nic Mach numbers greatly exceeding 2.5 cause the clumps
that they hit to evolve on too short of a timescale compared to the
lifetimes of observed molecular clouds. Significantly weaker shocks do
not lead to the creation of magnetically dominated regions.

\section{A Uniform, Static Medium with Cosmic Rays}

Using a two-fluid approach like that employed by
\citet{McKenzie:1982}, \citet{Wagner:2005} examined the stability of a perturbed
uniform radiative medium having a uniform cosmic ray
pressure. \citet{Shadmehri:2009} 
performed a similar analysis but also included a large-scale
magnetic field and thermal conduction, which we neglect here. Additional
relevant parameters include:  

\begin{eqnarray}
&z\equiv\frac{\omega}{ak}\\
%&a\equiv\left(\frac{\gamma_gP_{g0}}{\rho_0}\right)^{\frac{1}{2}}\\
%&k_T\equiv\frac{\mu(\gamma_g-1)}{Ra}\left.\frac{\partial\mathcal{L}}{\partial{}
%&k_\rho\equiv\frac{\mu(\gamma_g-1)\rho_0}{RaT_0}\left.\frac{\partial\mathcal{L}
&k_c\equiv\frac{a}{\chi}\\
&\phi\equiv\frac{\gamma_cP_{c0}}{\gamma_gP_{g0}}.
\end{eqnarray}

\noindent $\omega$, $k$, $P_c$, $\gamma_c$, and $\chi$ are
the angular frequency of the perturbation, its wavenumber, the cosmic ray
pressure, the adiabatic index of the cosmic ray fluid, and the cosmic ray
diffusion coefficient, respectively.

The dispersion relation is

\begin{eqnarray}\label{dispersion}
&G(z)\equiv{}z^4-iz^3\left(\frac{k}{k_c}+\frac{k_T}{k}\right)-z^2\left(\frac{k_T}{k_c}+
\phi+1\right)+ \nonumber \\
&iz\left(\frac{k_T}{k}\phi+\frac{k}{k_c}+\frac{k_T-k_\rho}{\gamma_gk}\right)+
\frac{k_T-k_\rho}{\gamma_gk_c}=0.
\end{eqnarray}

Stability requires that

\begin{eqnarray}
&\frac{k_T-k_\rho}{\gamma_gk_c}>0\label{ca}
\end{eqnarray}

\noindent as well as

\begin{eqnarray}
&\frac{k_T\phi+\frac{k_T-k_\rho}{\gamma_g}+\frac{k^2}{k_c}}{k_T+
\frac{k^2}{k_c}}>\frac{1}{2}\left(\phi+\frac{k_T}{k_c}+1\right)- \nonumber \\
&\sqrt{\frac{1}{4}\left(\phi+\frac{k_T}{k_c}+1\right)^2-
\frac{k_T-k_\rho}{\gamma_gk_c}}\label{cb}
\end{eqnarray}

\noindent and

\begin{eqnarray}
&\frac{k_T\phi+\frac{k_T-k_\rho}{\gamma_g}+\frac{k^2}{k_c}}{k_T+ 
\frac{k^2}{k_c}}
<\frac{1}{2}\left(\phi+\frac{k_T}{k_c}+1\right)+ \nonumber \\
&\sqrt{\frac{1}{4}\left(\phi+ \frac{k_T}{k_c}+1\right)^2-
\frac{k_T-k_\rho}{\gamma_gk_c}}\label{cc}.
\end{eqnarray}

\noindent The first of these conditions implies that cosmic rays do
not influence the thermal stability of a medium with respect to isobaric
perturbations,
though cosmic rays can affect the growth rate. The other conditions
imply that cosmic rays do influence the thermal stability of a medium
with respect to isentropic perturbations. The final two conditions are
sufficiently complicated that examination of them in various limits
is desirable. For instance, in the limit of very large $\phi$ and very
large $k_c$, the growth of insentropic perturbations is negligible
for all cooling functions.  

\section{Radiative Shocks without Cosmic Rays}

\citet{Falle:1975} and \citet{McCray:1975} first showed that radiative shocks are
in some cases unstable. Later \citet{Falle:1981} and \citet{Langer:1981} discovered that under some
circumstances they oscillate due to global overstability. Numerical work \citep[e.g.,][]{Imamura:1984, Strickland:1995} showed that high Mach number
shocks are overstable if $\alpha < 0.4$, in good agreement with the results
of a linear stability ananlysis
\citep{Chevalier:1984}. \citet{Pittard:2005} 
showed that the range of values of $\alpha$ for which radiative
shocks are overstable depends on the ratio of the
far downstream temperature to the far upstream temperture and the Mach
number. For a fixed value of that ratio of  unity, the maximum value of
$\alpha$ for which a shock is overstable decreases with decreasing
Mach number.

Though the purely hydrodynamic models indicate that many radiative shocks
in supernova remnants should be non-steady, steady shock models have
typically been used with reasonable success in the analysis of observations
of radiative supernova remnant shocks \citep[e.g.,][]{Raymond:2001}.

\section{Radiative Shocks with Cosmic Rays}

\citet{Wagner:2006} adopted the two-fluid approach to constructing
plane-parallel models of radiative
shocks to determine the effects of cosmic rays on the development of
the overstability. They assumed that the magnetic field is dynamically
unimportant and that the only energy transfer from the
cosmic ray fluid to the thermal fluid occurs due to the inclusion of the
derivative of the cosmic ray pressure in the equation governing the momentum
of the thermal fluid. In most models they ignored the effects of cosmic
rays long enough for the overstability to develop as it would in a single
radiative thermal fluid. Then an initially uniform cosmic ray pressure was
introduced. The distant upstream value of this pressure was maintained at
a constant value, and the evolution of the pressure elsewhere was governed
by the appropriate moment of the the transport equation.

\citet{Wagner:2006} found that, no matter what the distant upstream cosmic
ray pressure is, the downstream flow behind a shock with a Mach number,
$M \equiv V_s/a$ where $V_s$ is the shock speed, of 3 or more
is cosmic-ray dominated. Cosmic-ray dominated shocks are stable
and steady. Their model implies that shocks that are radiative rather
than adiabatic are cosmic-ray dominated for a wider range of Mach number,
preshock cosmic ray pressure parameter space. For sufficiently large
ratios of the diffusion length to the cooling length, the cosmic-ray
dominated shocks are nearly isothermal with structures similar to those
given by analytic solutions for cosmic-ray dominated, strictly isothermal
flows. 

The 2006 model is
not realistic. It implies that the postshock to preshock density ratio
is close to 7 in shocks with $M$ of 3 or more, a result that is not
consistent with previous efforts to interpret observations of radiative
supernova remnant shocks \citep[e.g.,][]{Raymond:2001}.

Greater compression factors can be obtained for models in which
sufficient cosmic ray energy is transformed into thermal energy of the
thermal fluid. One mechanism causing the required transformation is the
acoustic instability that \citet{Drury:1986} found to occur in shocks when
$\chi/a$ exceeds the absolute value of $ \gamma_c P_c
(\partial{P_c}/\partial{x})^{-1}$, where the $x$-axis is parallel to the shock
velocity. To account for heating due to the acoustic instability, \citet{Wagner:2007} included a source term in
the thermal fluid energy equation. When the acoustic instability condition
is not met, the term is zero. When the condition is strongly satisfied,
the term is approximately $3P_g/2\tau$ for $\gamma_g = 5/3$, where
$\tau$ is a specified time constant. When the condition is very
weakly satisfied, the term approaches 0.

\citet{Wagner:2007} showed that if $V_s \tau$ is small enough compared to
the diffusion length and the cooling length, shocks with Mach numbers of
5 and 10 behave very similarly to their single fluid counterparts and
display thermal overstability for appropriate radiative cooling laws. For
strong shocks one would hope to be able to tune the ratios
of the energy transfer length, diffusion length, and cooling length to
obtain structures showing a continuous range of behaviour from steady
cosmic-ray dominated flows to overstable flows with small cosmic ray
pressure everywhere. Though we know that values of the ratios can
be selected to give cosmic-ray dominated flow or flow resembling that
in a single-fluid radiative shock, we do not know that a smooth transition
from one type of extreme behavior to the other occurs in general when the
ratios vary smoothly. Indeed, attempts made by \citet{Wagner:2007} to
find a combination of parameters leading to a strong shock solution
with a moderate postshock ratio of the cosmic ray and thermal pressures
were unsuccessful. \citet{Becker:2001} have performed a thorough
analysis of the very closely related bifurcation of solution space
for adiabatic cosmic-ray modified shocks.  

\section{Conclusions}

Theoretical studies are interesting in themselves, but ultimately one aims
to use models to understand observations. So far, the \citet{Wagner:2007}
models of cosmic-ray moderated radiative shocks have not been used to
interpret data.

However, \citet{Boulares:1988} attempted to use similar models of
adiabatic cosmic-ray modified shocks to interpret optical data for
the Cygnus Loop. They did not include energy transfer from the cosmic
rays to the thermal fluid due to the \citet{Drury:1986} acoustic
instability, and optical data have improved sufficiently that more reliable
inferences are now possible.

\citet{Wagner:2009} used time-dependent adiabatic shock models,
incorporating energy transfer from
the cosmic rays to the thermal fluid in the manner
described above, to interpret optical emission data for knot g in the Tycho
supernova remnant. They inferred values for the diffusion coefficient,
the injection parameter, and the energy timescale of $2 \times 10^{24}$
cm$^2$ s$^{-1}$, $4.2 \times 10^{-3}$, and 426 yr, respectively.

\begin{acknowledgements}

This work has been supported by the UK Science and Technology Facilities
Council and the Royal Society.

\end{acknowledgements}

\bibliographystyle{aa}

\end{document}